\documentclass[%
 aip,
 amsmath,amssymb,
 reprint,%
]{revtex4-1}

\usepackage{graphicx}
\usepackage{dcolumn}
\usepackage{bm}
\usepackage{hyperref}
\usepackage{amsmath}
\usepackage{color}
\usepackage{physics}
\usepackage[dvipsnames]{xcolor}
\usepackage[normalem]{ulem}

\definecolor{newtext}{RGB}{0, 0, 0}

\begin{document}

\preprint{}

\title{Tunable quasi-discrete spectrum of spin waves excited by periodic laser patterns}

\author{Ia. A. Filatov}
 \email{yaroslav.filatov@mail.ioffe.ru}
 \homepage{https://ioffe.ru/ferrolab/}
\author{P. I. Gerevenkov}
\author{N. E. Khokhlov}
\author{A. M. Kalashnikova}

\affiliation{%
    Ioffe Institite, 26 Politekhnicheskaya, 194021, St. Petersburg, Russia
}%

\date{\today}

\begin{abstract}
We present a concept for selective excitation of magnetostatic surface waves with quasi-discrete spectrum using spatially patterned femtosecond laser pulses inducing either ultrafast change of magnetic anisotropy or inverse Faraday effect.
We micromagnetically simulate excitation of the waves with periodically patterned uni- or bipolar laser impact.
Such excitation yields multiple wavepackets propagating with different group velocities, whose dispersion corresponds to the set of quasi-discrete points.
In addition, we show that the frequency of the spectral peaks can be controlled by the polarity of the periodic impact and its spatial period.
Presented consideration of multiple spatially periodic \textcolor{newtext}{magnetostatic surface wave} sources as a whole enables implementation of a comprehensive toolkit of spatio-temporal optical methods for tunable excitation and control of \textcolor{newtext}{spin wave} parameters.
\end{abstract}


\maketitle


\section{\label{sec:intro} Introduction}

Recent progress in magnonics demonstrates multiple ways to use spin waves (SW) as data carriers for non-charge-based energy-efficient computation \cite{MagnonicsRoadmap24}.
Implementing SW offers wide possibilities for processing information, when conventional Boolean operations are realized in magnonic integrated circuits~\cite{mahmoud2022would, wang2020magnonic, garlando2023numerical}.
This promotes THz operation rates as well as scaling down the devices to the nanometer range comparable to the wavelengths of SW~\cite{wang2023deeply, hortensius2021}. 
Furthermore, the lack of Ohmic losses during SW propagation, the wide available material range, and the potential to employ non-boolean wave-based calculations make magnonics a highly promising base for the new-generation computing~\cite{mahmoud2020introduction, papp2021nanoscale, yaremkevich2023chip}.

The use of coplanar antennas for excitation and detection of SW as well and its combination with Brillouin light scattering is now considered as a standard in magnonics.
This approach revealed a variety of effects of SW interaction to be implemented in logic devices, when SW is excited continuously~\cite{talmelli2020reconfigurable,wang2020magnonic,sadovnikov2015magnonic,sadovnikov2022exceptional} or by nanosecond microwave pulses~\cite{breitbach2024nonlinear,mahmoud2022would}.
However, achieving of the highest performance requires an excitation of SW wavepackets of sufficiently short duration, which may increase the clock speed of information processing.
Femtosecond laser pulses have been utilized for controllable excitation of such SW wavepackets since the first demonstrations \cite{Satoh2012,van2002all,au2013direct,yoshimine2014phase}.
Further exploration of various means of active control of parameters of laser-induced SW, such as phase and group velocities, is required for developing optomagnonic concepts~\cite{bublikov2023laser,kolosvetov2022concept}.

Optically reconfigurable magnonics uses optically-induced changes of material parameters to tune SW properties~\cite{vogel2015optically}.
Combined with femtosecond laser pulses, it offers comprehensive tools for SW investigation in the spatial, temporal, and frequency domains with sub-micrometer and sub-picosecond resolutions~\cite{chernov2020all,hortensius2021,hashimoto2017all,khokhlov2019optical}.
Moreover, optical methods allow one to shape excitation in a spatially non-trivial form, and thus tune wavenumber distribution of SW \cite{Satoh2012}.
Recently, the interaction of laser-induced SW from a pair of localized sources was demonstrated~\cite{Yoshimine_EPL2017, kolosvetov2022concept, Kolosvetov_IEEE2023}.
The development of a model describing composite spatially patterned laser excitation as a whole and the analysis of associated magnetization dynamics both inside and outside the excitation pattern can simplify its practical implementation and can reveal additional ways to control SW parameters~\cite{talmelli2020reconfigurable}.

In this article, we develop the concept of selective excitation of magnetostatic surface waves (MSSW) with quasi-discrete spectrum using spatially patterned femtosecond laser pulses.
The proposed spatial distributions of the intensity and polarization of excitation pulses can be obtained experimentally using a spatial light modulator, as used, for example, in Refs.~\cite{kolosvetov2022concept, Kolosvetov_IEEE2023}.
We consider a model of MSSW excitation due to the two experimentally verified mechanisms: ultrafast thermal change of magnetic anisotropy (CMA) and inverse Faraday effect (IFE), and we validate this model in the case of single pulse excitation using micromagetic simulations.
Analyzing the dispersion of thus excited MSSW, we show no significant difference between waves propagating beyond the excitation area.
Second, we micromagnetically simulate excitation of MSSW with periodically patterned laser pulses.
Periodic excitation via IFE is considered to be either uni- or bipolar, depending on the modulation of the helicity of the excitation pattern, while CMA excitation is unipolar only.
Considering pump pulses periodically distributed in space as a single SW source, we show that it has a quasi-discrete periodic spectrum in reciprocal space with either even or odd modes.
Indeed, MSSW characteristics outside the excitation area are governed by the polarity and periodicity of the excitation pattern.
Inside the pattern, the excitation mechanism dictates the spin wave evolution. 
Thus, upon long-lived CMA effect, periodicity of the pump in space leads to the formation of a transient reconfigurable magnonic crystal structure~\cite{vogel2015optically}.

\section{Model of MSSW excited by periodic laser pattern}\label{Sec:Model}

\subsection{Geometry of MSSW laser excitation}

\begin{figure}
    \centering
    \includegraphics{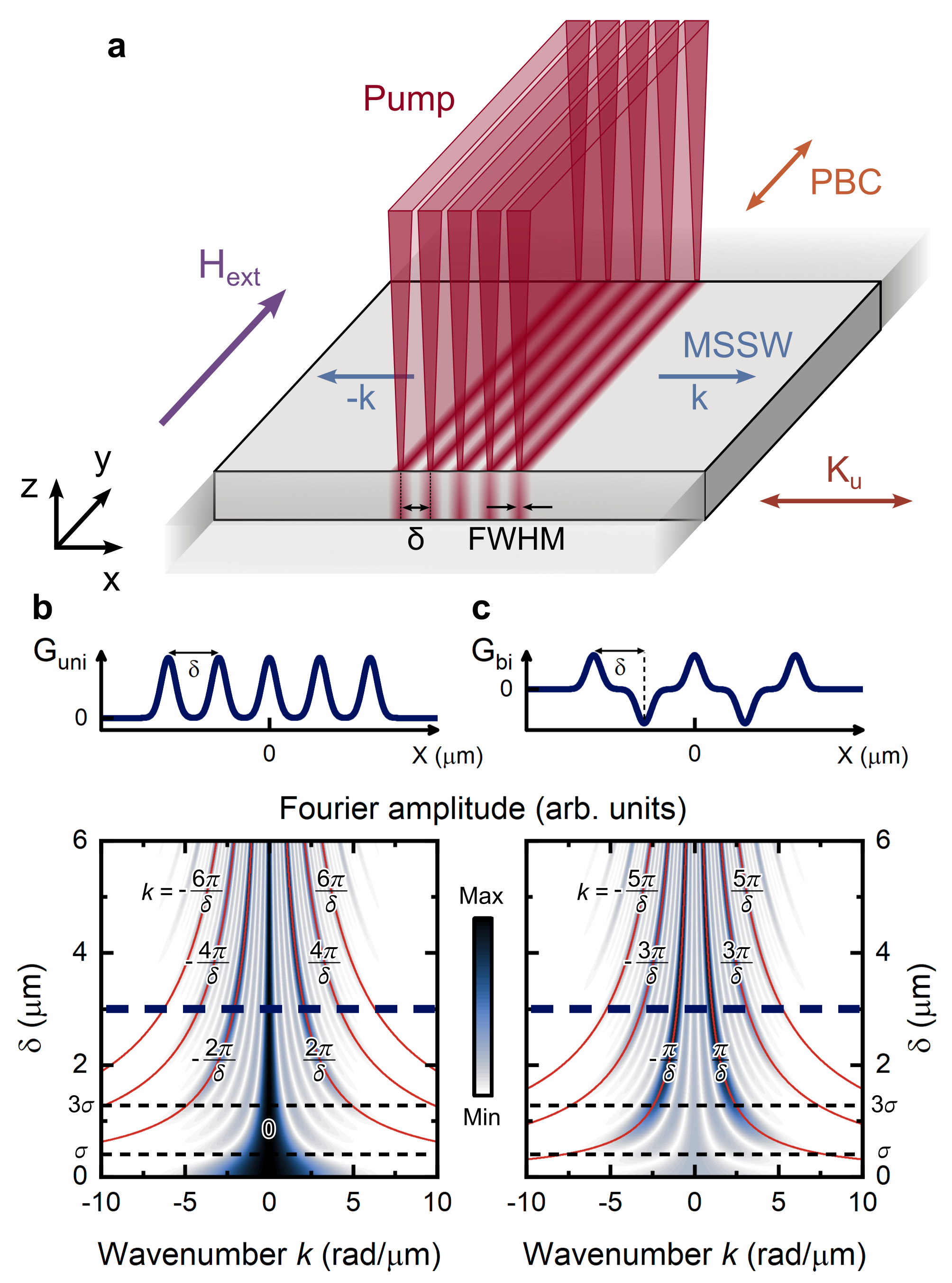}
    \caption{\label{fig:geometry}
    \textbf{The geometry of the plane MSSW excitation with periodic pump pattern and the corresponding spectral distribution.}
    \textbf{a}, the geometry under investigation.
    \textbf{b}, \textbf{c}, the distributions of the Fourier amplitudes calculated analytically using \eqref{FourierFull} for uni- and bipolar impact, respectively. The spatial profiles are illustrated above the corresponding maps. The numbers in white denote the even and odd order of the modes with corresponding values of wavenumber $k$ indicated by solid red lines. Horizontal black dashed lines illustrate the values $\sigma$ and $3\sigma$. The horizontal blue dashed line corresponds to $\delta=3$\,$\mu$m used in the following simulations.
    }
\end{figure}

To construct the model of spatially periodic laser excitation of MSSW, we consider bismuth-doped yttrium iron garnet Bi:YIG as a magnetic material, since it supports CMA and IFE for excitation of magnetization dynamics \cite{Shelukhin2018}.
As we are interested only in the excitation of MSSW, we reduce the model to one-dimensional (1D) geometry (Fig.\,\ref{fig:geometry}\,a), and consider a film, where plane MSSW are excited by a laser spot limited in the $x$ direction and infinitely long in the $y$ direction, as in experiments with elongated pump laser spots~\cite{Satoh2012, kainuma2021fast, matsumoto2020observation, hioki2020bi, hioki2022coherent, Yoshimine_EPL2017}.
\textcolor{newtext}{This also excludes an excitation of backward volume magnetostatic spin waves with nonzero wavenumber $k$.}
In-plane external magnetic field $\textbf{H}_{\text{ext}}$ is oriented along the $y$ axis, and MSSW propagate perpendicularly to $\textbf{H}_{\text{ext}}$, i.e. along $x$ axis.
In our model, the uniaxial magnetic anisotropy axis with the equilibrium parameter $K_u$ is oriented along $x$ axis to fulfill the condition of optical excitation of MSSW due to CMA \cite{khokhlov2019optical, filatov2020spectrum}.
Also, we assume the excitation to be uniform across the thickness of the film to exclude the \textcolor{newtext}{formation of perpendicular} standing spin waves (\textcolor{newtext}{P}SSW).

To describe periodically patterned excitation of MSSW due to either CMA and IFE, we introduce the spatio-temporal profile of laser impact as:
\begin{equation}\label{generalImpact}
    I(x,t)=I_{0} G(x) T(t),
\end{equation}
where $I_{0}$ corresponds to the amplitude of the impact; $G(x)$ and $T(t)$ describe spatial and temporal profiles of the impact, respectively.
We assume that the individual pump spot has a Gaussian profile in space.
Thus, the total spatially periodic excitation profile is determined by 1D function of $2N+1$ Gaussians centered at $x=0$ and periodically spaced by a distance $\delta$ (upper panel in Fig.\,\ref{fig:geometry}\,b):
\begin{equation}\label{GspaceUNI}
    G_{\text{uni}}(x,N,\delta)=\sum_{n=-N}^{N} \exp\left[ -\frac{(x + n \delta)^2}{2\sigma^{2}} \right],
\end{equation}
where $\sigma=\text{FWHM}/(2\sqrt{2\ln{2}})$, and full width at half maximum (FWHM) is fixed to 1\,$\mu m$ being larger than the diffraction limit to fulfill experimental conditions~\cite{khokhlov2019optical, filatov2022spectrum, iihama2016quantification}.
Equation \eqref{GspaceUNI} describes a single Gaussian function if $N=0$, and adds $\pm N$ spatially periodic Gaussian functions located on both sides of the central one with separation by a period $\delta$.
Such unipolar pattern is applicable for both CMA excitation and IFE excitation with a fixed circular polarization of laser pulses
To account for the bipolarity of IFE when the polarization of the laser pulses is altered between the neighbouring spots \cite{kimel2005ultrafast}, Eq.\,\eqref{GspaceUNI} is modified as (upper panel in Fig.\,\ref{fig:geometry}\,c):
\begin{equation}\label{GspaceBI}
    G_{\text{bi}}(x,N,\delta)=\sum_{n=-N}^{N} (-1)^n \exp\left[ -\frac{(x + n \delta)^2}{2\sigma^{2}} \right].
\end{equation}

The temporal profile $T(t)$ of the impact Eq. \eqref{generalImpact} depends on the physical process induced by the laser in the material, as discussed below.

\subsection{Physical impacts of the laser pulse}

Laser-induced CMA is introduced as an ultrafast reduction of the equilibrium uniaxial magnetic anisotropy parameter $K_u$ with the maximal relative amplitude $\Delta K_u$.
In dielectric magnets, the CMA occurs on the picosecond time scale~\cite{Shelukhin2018}, which is 2-3 orders of magnitude faster than the MSSW period~\cite{chernov2017optical}. 
In turn, relaxation processes are on the nanosecond time scale, which is comparable to the typical propagation time of magnetostatic waves~\cite{Satoh2012, matsumoto2020observation, kainuma2021fast}.
Thus, we approximated the temporal dependence of laser-induced CMA by the Heaviside step function $T_H(t)=\Theta(t-t_0)$, where $t_0$ is the moment of pump impact.
The resulting spatio-temporal profile of periodically patterned CMA is:
\begin{equation}\label{Ku}
    K_u'(x,t,N,\delta) = K_u\left[1-\Delta K_u G_{\text{uni}}(x,N,\delta) T_H(t) \right].
\end{equation}

The laser-induced IFE in our model is introduced as a pulse of the effective out-of-plane field $\textbf{H}_{\text{IFE}}$, according to experimental observations~\cite{kimel2005ultrafast, kirilyuk2010ultrafast}.
The field follows the temporal envelope of femtosecond laser pulse and is described by the Gaussian profile \mbox{$T_G(t)=\exp\left[-(t-t_0)^{2}/\left(2\sigma_t^2\right) \right]$}, where \mbox{$\sigma_t=\text{FWHM}_t/(2\sqrt{2\ln{2}})$} and FWHM$_t$ is set to 200 fs to meet the typical experimental pump pulse duration~\cite{yaremkevich2021protected,khokhlov2019optical,hioki2020bi,krichevsky2024spatially}.
In addition, we take into account the possibility to control the polarity of $\textbf{H}_{\text{IFE}}$ by the choice of right- or left-handed circular polarization of pump pulses in experiments~\cite{kimel2005ultrafast, kirilyuk2010ultrafast}.
Thus, the periodically patterned IFE in our model is represented by unipolar~\eqref{GspaceUNI} or bipolar~\eqref{GspaceBI} impact in space.
The resulting spatio-temporal profile of the IFE with maximal amplitude $H_{\text{IFE}}$ takes the form:
\begin{equation}\label{uniIFE}
    H_{\text{IFE}}'(x,t,N,\delta) = H_{\text{IFE}} G(x,N,\delta) T_G(t),
\end{equation}
where $G(x,N,\delta)$ has the form \eqref{GspaceUNI} or \eqref{GspaceBI}.

Note that the approximation ~\eqref{Ku} works for thermally induced CMA, which only has the unipolar impact~\eqref{GspaceUNI}.
However, there are other mechanisms of CMA in iron garnets, including non-thermal ones \cite{kirilyuk2010ultrafast, stupakiewicz2017ultrafast} with alternating sign.
Laser-induced demagnetization of a ferromagnet usually appears along with CMA \cite{Gerevenkov2021,Maehrlein2018}.
However, it has much longer rise times in iron garnets and has a value of an order of magnitude smaller than CMA~\cite{Shelukhin2018,Davies2019,Deb2021}.
Thus, we omitted the demagnetization process in the model.

The approximation of the excitation being homogeneous across the thickness is valid for iron garnets with thicknesses ranging from tenths to hundreds of nanometers.
In general, however, pump pulse impact in iron garnet films is non-uniform across the thickness in accordance with light penetration depth\textcolor{newtext}{, which has typical values of about 100\,nm for photons with energy above absorption edge~\cite{Wemple1974} used for CMA and two-three orders of magnitude larger for photons below absorption edge used for IFE.}
This should cause the excitation of \textcolor{newtext}{P}SSW.
We omit this from the consideration in order to focus on propagation of MSSW. 
We note that high-quality ultrathin garnet films where our assumption is valid are now becoming available \cite{Levy:19,Hartmann2024}.

\subsection{\label{subsec:k_distr} Analysis of wavenumber distribution}

To analyze the excitation of the MSSW via CMA and IFE, we investigate the distribution of the wavenumbers $k$, excited by such a periodically patterned impacts.
The wavenumber distribution of the laser-induced SW amplitudes is known to be determined by the Fourier transform of the pump spatial profile \cite{Satoh2012, hortensius2021}.
We find analytically the Fourier amplitudes as a function of wavenumber for uni- and bipolar spatially periodic patterns (\ref{GspaceUNI},\ref{GspaceBI}).
Then, the Fourier transform for arbitrary $N$ has the form:
\begin{equation}\label{FourierFull}
    \widetilde{A}(k, N, \delta) = \sqrt{   \frac{\cos{ \left[k \delta (2N+1) \right]} \mp 1}{\cos{ \left(k \delta \right)} \mp 1}   } \sigma \exp\left(-\frac{k^2\sigma^2}{2}\right),
\end{equation}
where top and bottom signs correspond to the uni- and bi-polar excitation, respectively.
To demonstrate the main findings, we limited the total number of Gaussian functions to 5 ($N=2$) in further consideration.
In Eq.~\eqref{FourierFull}, the first factor is a Dirac comb~\cite{kindermann2011landau} for finite $N$. 
The spatially limited Dirac comb we propose also demonstrates the same periodicity, but allows us to consider wave propagation both inside the pattern and outside it.

We calculate the corresponding Fouirer amplitude distribution of the modes as the dependence on the spatial period of the pump pattern $\delta$ (Fig.\,\ref{fig:geometry}\,b and c).
In the case of unipolar impact, the distribution has maximal amplitudes around even order modes starting at $k=0$ (Fig.\,\ref{fig:geometry}\,b).
In contrast, the bipolar impact leads to emergence of odd order modes (Fig.\,\ref{fig:geometry}\,c).
This distribution is multiplied by the overall decrease in amplitude with $k$ governed by the spatial width $\sigma$ of the individual laser spot in the pattern and described by the exponential factor in Eq.~\eqref{FourierFull}.
The spacing between the spots $\delta$ controls both the wavevnumbers and the amplitudes of the excited modes, as seen in Fig.\,\ref{fig:geometry}\,b and c.
For example, increasing $\delta$ leads to excitation of higher-order modes, as their wavenumbers decrease and they fall within a range of $k$ limited by the size of the single spot.
Thus, while the total width of the excited wavenumber is determined by \textcolor{newtext}{$\sigma$~\cite{Satoh2012,jackl2017magnon}}, the periodicity establishes the quasi-discrete set of wavenumbers with spectral positions controlled by the spatial period $\delta$.
As expected, the quasi-discrete spectrum emerges if $\delta$ exceeds $3\sigma$, and the excitation pattern acquires a contrast higher than $30\%$.
We also note that the change of the total number of Gaussians in periodic pattern affects their spectral width (see Appendix~\ref{sec:appendixNgauss}).

The established model outlines a possibility to control SW parameters using periodically patterned laser excitation in addition to the tunability of SW laser excitation based on pump spot shape and size~\cite{Satoh2012, chernov2017optical}, mutual orientation of the external magnetic field and magnetic anisotropy axes \cite{khokhlov2019optical, Khokhlov2024}.
Additional tools are a spatial period of the pattern $\delta$, a choice of excitation mechanism, CMA or IFE, as well as a choice between uni- and bipolar excitation as seen from Eq.\eqref{FourierFull} and Fig.\,\ref{fig:geometry}\,b and c. 
According to our model, CMA being always unipolar impact offers an additional way to control SW parameters.
In particular, CMA is a long-living effect compared to MSSW propagation, which results in the formation of periodic non-uniformity of $K_u$ in space.
Thus, one expects that the part of excited MSSW can stay inside this structure, filling it as a magnon resonator.
Further, we support the above analytical predictions with the results of micromagnetic modeling.

\begin{figure*}
    \includegraphics{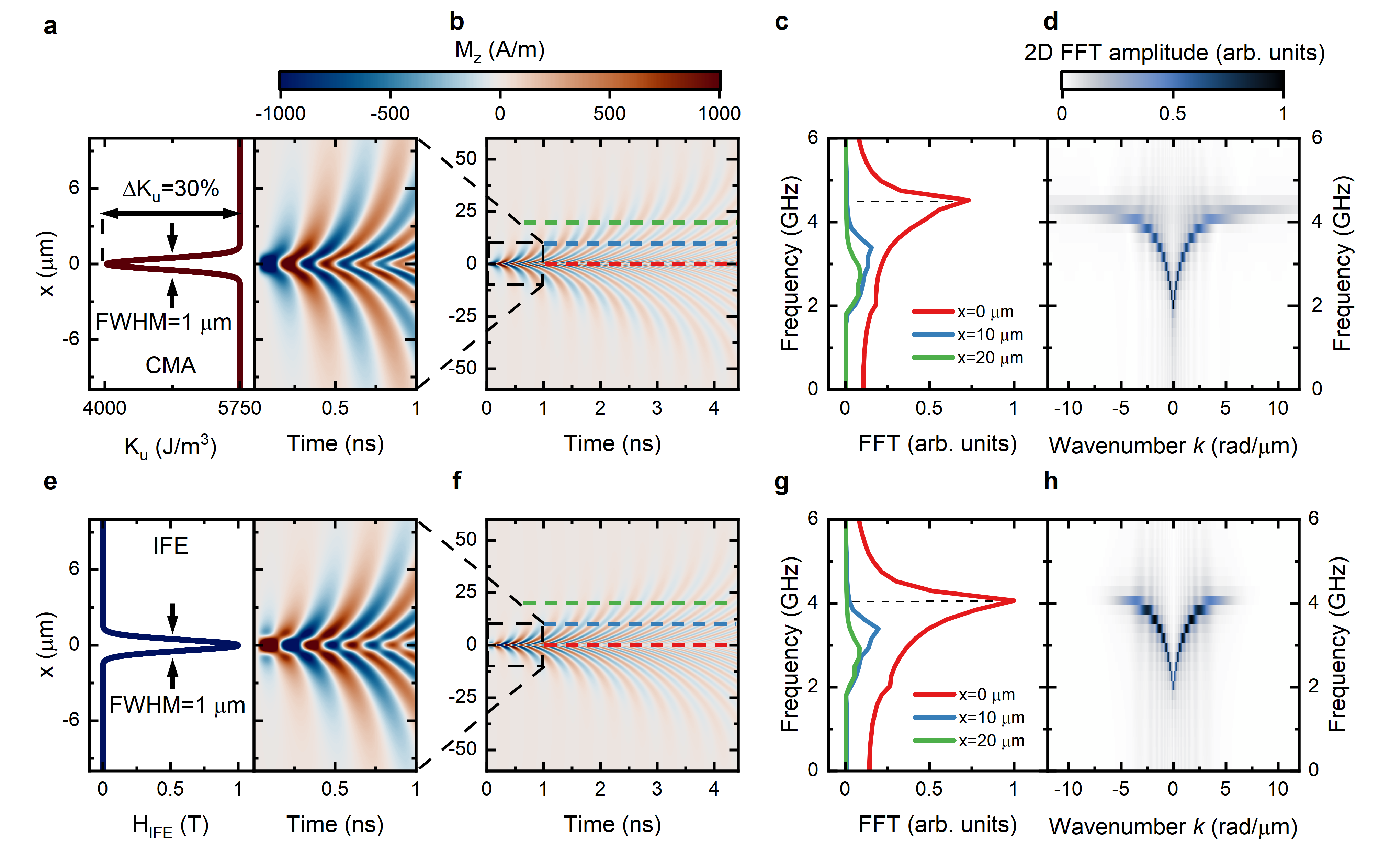}
    \caption{\textbf{MSSW excitation by the single spatial Gaussian impact.} 
    The panels \textbf{a-d} and \textbf{e-h} correspond to the excitation by the CMA and IFE, respectively.
    \textbf{a}, \textbf{e}, Spatial profiles of the excitation mechanisms (left), and the spatio-temporal distribution of the out-of-plane magnetization component $M_z(t,x)$ in the vicinity of the excitation area (right).
    \textbf{b}, \textbf{f}, $M_z(t,x)$ dynamics throughout the entire range of time and distance.
    \textbf{c}, \textbf{g}, Frequency spectra of magnetization dynamics at various distances from the excitation area obtained by 1D FFT of $M_z(t)$ at fixed coordinates $x=$ 0, 10, 20\,$\mu$m depicted with dashed lines of corresponding color in panels \textbf{b} and \textbf{f}.
    The black dashed lines show the high-frequency maximum.
    \textbf{d}, \textbf{h}, Excited wave dispersions obtained by 2D FFT of the $M_z(t,x)$ data in panels \textbf{b} and \textbf{f}.
    }
    \label{fig:singlepump}
\end{figure*}

\section{Micromagnetic simulations}

\subsection{Modeling parameters}\label{subsec:modelParameters}

Micromagnetic validation of the proposed model is performed using Ubermag \cite{Beg2022} with object oriented micromagnetic framework (OOMMF) \cite{OOMMF} as a micromagnetic calculator.
In simulations, we use the material parameters, typical for Bi:YIG films~\cite{zeuschner2022standing}: magnetization saturation $\mu_0 M_s=175$\,mT, exchange energy constant $A_{ex}=3.5\times10^{-12}$\,J/m, Gilbert damping $\alpha=0.01$, and uniaxial anisotropy parameter $K_u=5750$\,J/m$^3$.
The in-plane external magnetic field $\mu_0 H_{\text{ext}}=100$\,mT is applied with a tilt of 1$^{\circ}$ with respect to \textcolor{newtext}{$y$} axis.
We set the lateral size of the computational cell as $10\times10$\,nm$^2$ to be smaller than the magnetostatic exchange \textcolor{newtext}{length} $\sqrt{2A/(\mu _0M_s^2)} = 16.9$\,nm and the magnetocrystalline exchange length $\sqrt{A/K_u} = 24.7$\,nm~\cite{abo_definition_Lex_IEEE_2013}.
The thickness of the film is 500\,nm.
The corresponding sagittal cell size 500\,nm \textcolor{newtext}{allows excluding} the excitation of \textcolor{newtext}{P}SSW.
1D periodic boundary conditions are applied along \textcolor{newtext}{$y$} axis.
In the simulations, time-domain dynamics is tracked by solving the Landau-Lifshitz-Gilbert equation with the set of energy terms: Zeeman energy, exchange energy, energy of the uniaxial magnetic anisotropy, and magnetostatic energy.
CMA is simulated by introducing the spatially non-uniform parameter $\widetilde{K_u}(x,t,N,\delta)$ described by \eqref{Ku}.
IFE is simulated by adding to the system the Zeeman energy term with the transient field in the form \eqref{uniIFE}.
The \textcolor{newtext}{magnitudes} of CMA and IFE are close to the experimentally achievable values: $\Delta K_u=0.3$~\cite{carpene2010ultrafast,Gerevenkov2021} and $\mu_0 H_{\text{IFE}}=\pm 1$\,T~\cite{kimel2005ultrafast,kalashnikova2015ultrafast,kozhaev2018giant, Shelukhin2018}.
The ground state of the magnetization is calculated as a relaxed state after the initializing of the system with the equilibrium parameters, minimization of its energy, and time domain calculations during 10 ns with an increased Gilbert damping of $\alpha = 1$ without excitation.
\textcolor{newtext}{One total individual simulation including numerical Fourier analysis to get the dispersion relations took approximately 1.5 hours using 16-thread CPU at 5 GHz operating frequency.}

\subsection{Single pulse excitation}\label{Sec:single pulse}

First, we consider the excitation of MSSW using a single source, i.e., at $N = 0$ in Eq.~\eqref{GspaceUNI}.
In the spatio-temporal domain, it is common to monitor the out-of-plane magnetization component $M_z$, following the optical pump-probe experiments using Faraday~\cite{Satoh2012, chernov2017optical} 
or polar Kerr~\cite{khokhlov2019optical, filatov2022spectrum, Au_PRL2013} effects.
CMA and IFE are used as excitation mechanisms, and the corresponding spatial excitation profiles at $t = t_0$ are shown in the left panels of Fig.~\ref{fig:singlepump}\,a and e.
The right panels show the corresponding spatio-temporal dependencies of $M_z(t,x)$.
Within the excited area ($|x| < \sigma$), the laser-induced magnetization precession is observed at $t > t_0 = 25$\,ps.
The propagation of the MSSW is observed away from the excitation area in the form of a wavepacket.
Regardless of the excitation mechanism, the wave phase and group velocities of the same sign are characteristic for MSSW, and waves propagate over a distance of more than \textcolor{newtext}{23}\,$\mu$m within the modeled time window of 4.5\,ns (Fig.~\ref{fig:singlepump}\,b and f).
\textcolor{newtext}{The propagation lengths, \textit{i.e.} a length where amplitude decreases by a factor of $e$, varying between 2 and 23\,$\mu$m for high- and low-frequency components, respectively, are estimated by a standard exponential fit of fast Fourier transform (FFT) amplitudes of different spectral components.}
In addition, the initial phase of the waves is defined by the circular polarization direction in the case of IFE~\cite{kimel2005ultrafast, Satoh2012}.
\textcolor{newtext}{Note that excited magnetization dynamics is in a linear regime, and for the used CMA and IFE magnitudes nonlinear effects do not arise, as shown in Appendix~\ref{sec:appendixHIFEdep}.}

However, upon closer examination, the differences in the dynamics of $M_z(t,x)$ are observed between the two excitation mechanisms (right panels of Fig.~\ref{fig:singlepump}\,a and e):
a sharper phase front evolution around $x = 0$ in the IFE case indicates differences in frequencies between the two mechanisms.
To explain it, we consider the FFT of $M_z(t)$ traces at various distances from the excitation area (Fig.~\ref{fig:singlepump} c and g).
At $x = 0$ (red lines), an excitation of a wide frequency range is observed with two peacks near 2 and 4\,GHz for both IFE and CMA.
The low-frequency peak is more pronounced in the IFE case, and the high-frequency peak has a higher frequency in the CMA case (marked with a dashed line).
The wave dispersions for the two excitation mechanisms obtained by 2D FFT (Fig.~\ref{fig:singlepump} d and h) indicate that the two peaks correspond to the minimum and maximum frequencies of MSSW.
\textcolor{newtext}{This is consistent with the experimental data obtained with a round laser spot in Ref.~\cite{chernov2017optical} except for the absence of backward volume waves suppressed in our case due to the elongated excitation area shape.}
Out of the excitation area (blue and green lines in Fig.~\ref{fig:singlepump} c and g), the frequency spectra are similar for the two excitation methods and lie between the frequencies of the initial peaks.

\textcolor{newtext}{Let us discuss the differences between IFE and CMA in the excited spectra at $x = 0$\,$\mu$m.
We attribute them to the different time dependencies for the two excitation methods.}
In the case of IFE \eqref{uniIFE} the laser-induced field disappears in a time much shorter than the characteristic period of magnetization dynamics.
After pumping, the excited dynamics at $x = 0$ has a certain frequency spectrum and then MSSW propagates out of the excitation area with group velocities corresponding to its dispersion.
\textcolor{newtext}{Slow waves with frequency around 4\,GHz do not leave the excitation area and manifest themselves as a pronounced peak in FFT spectrum at $x=0$}.
In the case of CMA Eq.~\eqref{Ku} the material parameter $K_u$ experiences a long-lived partial quenching in the pump area.
\textcolor{newtext}{At magnetization along the hard anisotropy axis, the effective field is $| \mu_0 H_{\text{ext}} - 2K_u/M_s |$.
Decrease of $K_u$ leads to increase in the total effective field and hence to increase of the magnetization dynamics frequencies within the excitation area~\cite{filatov2022spectrum, filatov2020spectrum, khokhlov2019optical}. }
As a result, the state with minimum frequency \textcolor{newtext}{(ferromagnetic resonance)} inside the excited area corresponds to the states with non-zero group velocity outside this area.
\textcolor{newtext}{Thus, a low-frequency peak in the case of CMA is less pronounced (red line in Fig.~\ref{fig:singlepump}\,c).}
In contrast, the high-frequency part experiences a magnon "trap" due to CMA, since there are no magnon states at these frequencies outside the pumped area.
It leads to an increase in the maximum excited frequencies in the case of CMA and a more pronounced component with the wavenumbers $|k|>5$\,rad/$\mu$m observed inside the excitation area (Fig.~\ref{fig:singlepump}\,d).
Thus, the differences in the spectra are governed by the parts of the dispersion with near-zero group velocity and are most pronounced near the excitation area (see Appendix~\ref{sec:appendixSpectral} for details).
The propagating MSSW outside the pumped area can be considered identical for CMA and IFE.

\subsection{Periodically patterned excitation of MSSW}\label{sec:multipulse}

\begin{figure}
    \includegraphics{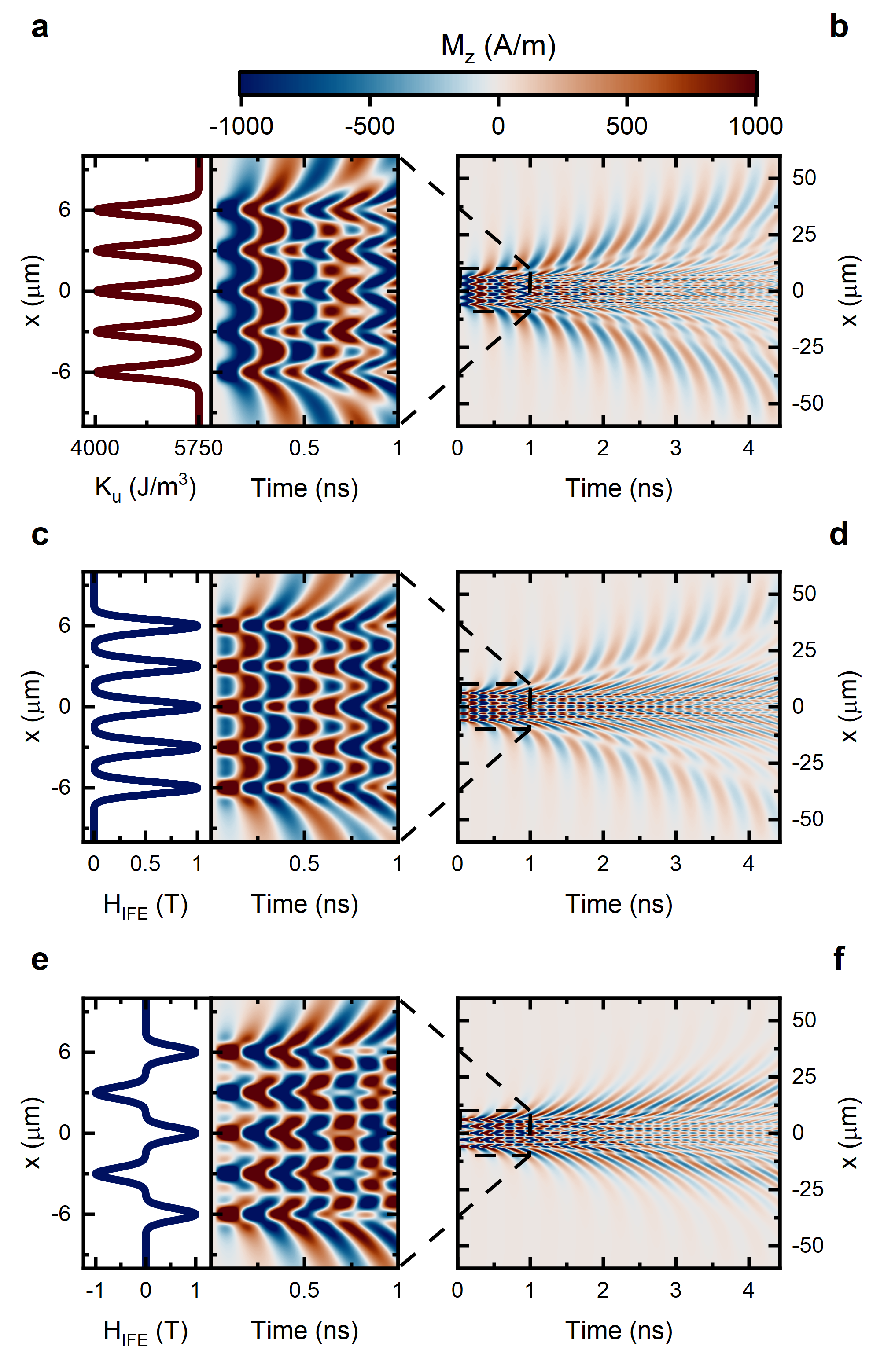}
    \caption{\textbf{MSSW excitation by spatially periodic Gaussian impacts with $2N+1=5$.} 
    The rows (\textbf{a,b}), (\textbf{c,d}) and (\textbf{e,f}) correspond to the excitation by  CMA, uni- and bipolar IFE, respectively.
    \textbf{a}, \textbf{c}, \textbf{e}, Spatial profiles of excitation mechanisms (left), and the spatio-temporal distribution of $M_z(t,x)$ in the vicinity of the excitation area (right).
    \textbf{b}, \textbf{d}, \textbf{f}, $M_z(t,x)$ dynamics throughout the entire range of time and distance.
    }
    \label{fig:Mz5pump}
\end{figure}

\begin{figure*}
    \includegraphics{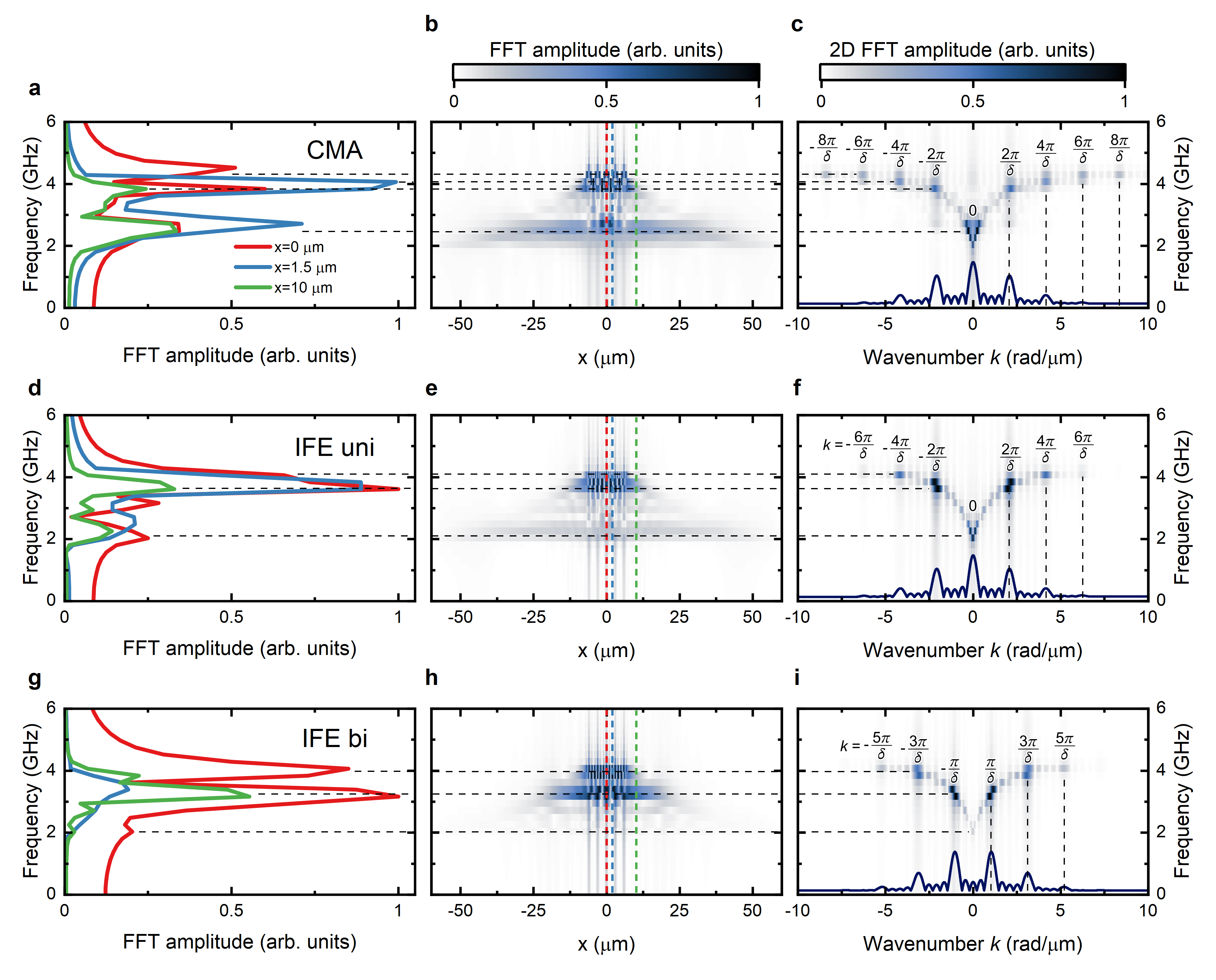}
    \caption{\textbf{Fourier analysis of MSSW excited by spatially periodic Gaussian impacts with $2N+1=5$.}
    The rows \textbf{a-c}, \textbf{d-f} and \textbf{g-i} correspond to the excitation by CMA, uni- and bipolar IFE, respectively.
    \textbf{a}, \textbf{d}, \textbf{g}, Frequency spectra of magnetization dynamics at various distances from the excitation area obtained by 1D FFT of $M_z(t)$ at fixed coordinates $x=$ 0, 1.5, 10\,$\mu$m.
    The coordinates are depicted with dashed lines of corresponding color in panels \textbf{b},\textbf{e} and \textbf{h}.
    \textbf{c}, \textbf{f}, \textbf{i}, Excited wave dispersions obtained by 2D FFT of $M_z(t,x)$ data in Fig.\,\ref{fig:Mz5pump}\,b,d and f, respectively.
    Horizontal black dashed lines in all panels indicate the same frequencies in the spectra.
    In the dispersions \textbf{c}, \textbf{f}, \textbf{i} orders of modes are specified and the corresponding wavenumber distributions calculated with \eqref{FourierFull} are shown by solid blue lines.
    }
    \label{fig:FFT5pump}
\end{figure*}

Next, we model the MSSW excitation via patterned CMA and uni- and bipolar IFE formed by 5 Gaussian pulses with a spatial period of $\delta=3$\,$\mu$m (Fig.\,\ref{fig:geometry}\,a).
Five spots in the pattern are sufficient to demonstrate the effect of periodicity~\cite{Qin2020yigMagnCryst}, and a further increase of $N$ does not qualitatively change the results (Appendix~\ref{sec:appendixNgauss}).
The chosen $\delta=3$\,$\mu$m is larger than $3\sigma=1.3$\,$\mu$m to ensure that there is no spatial overlap of the pumped areas.
The corresponding spatial profiles of laser-induced impact Eq.~(\ref{Ku}, \ref{uniIFE}) are shown in the left panels of Fig.\,\ref{fig:Mz5pump}\,a,c and e.
\textcolor{newtext}{We neglect a temporal evolution of the grating contrast and spatial profile on a timescale of our numerical simulations. 
To substantiate this, we estimate the temperature increase at 4\,ns after the excitation at the point $x=1.5$\,$\mu$m between two maxima. 
We use the heat conductivity~\cite{Euler2015} of 8\,W\,m$^{-1}$\,K$^{-1}$ and specific heat~\cite{Boona2014} of 600\,J\,kg$^{-1}$\,K$^{-1}$ for YIG. 
We assume the initial increase of the temperature of 10\,K at the center of the stripe ($x=0$), which is even somewhat higher than the typical temperature increase reported in literature~\cite{Shelukhin2018,zeuschner2022standing}, and consider the linear temperature gradient of 6.7$\cross$10$^{-6}$\,K\,m$^{-1}$ between the center of the stripe and the considered point. 
Under such conditions, the temperature increase at 4\,ns after the excitation at a point between two maxima is of $\approx$1\,K, showing that the pattern evolution due to lateral heat diffusion is just of 10$\%$. 
This rough estimate agrees with rigorous calculations of the temperature profile evolution in Bi:YIG~\cite{zeuschner2022standing}.
Thus, the negligible evolution of the grating profile is an appropriate approximation.}

As in the case of single spot excitation discussed above, magnetization dynamics possesses distinct features within the area subjected to the laser pattern and outside this area depending on the excitation mechanism. 
First, comparing $M_z(t,x)$ within the excitation area, we observe a pronounced difference of phase fronts depending both on the excitation mechanism, CMA or unipolar IFE, and on the polarity of IFE (right panels in Fig.\,\ref{fig:Mz5pump}\,a,c and e).
In all three cases, right after the excitation, the phase front is periodic in space and follows the excitation pattern. 
In the case of IFE, such a spatially periodic phase is also observed at longer time delays because the material preserves its homogeneous magnetic characteristics. 
The apparent change of the phase front at longer delay times is a result of constructive and destructive interference of the MSSW.
When CMA excitation is considered, there is an evident evolution of the phase front already after 1 temporal period. 
As seen in the right panel of Fig.\,\ref{fig:Mz5pump}\,a, the phase front within the excitation spots and between them becomes inhomogeneous. 
Furthermore, the temporal period of the oscillations of $M_z$ changes with time (see the temporal evolution at $x=0,\pm 3,\pm 6$\,$\mu$m).
This intricate behavior can be understood as an evolution of the spin-wave distribution inside a set of magnon resonators, which eventually results in a establishing of MSSW dispersion characteristic to a magnonic crystal.

The distribution of $M_z(t,x)$ throughout the spatio-temporal area demonstrates the propagation of multiple MSSW wavepackets with different group velocities (Fig.\,\ref{fig:Mz5pump}\,b,d and f).
The group velocities of the wavepackets are approximately the same when excited by CMA and unipolar IFE (Fig.\,\ref{fig:Mz5pump}\,b and d).
The change in IFE polarity results in the excitation of MSSW wavepackets with decreased group velocities (Fig.\,\ref{fig:Mz5pump}\,d and f).
Although the temporal period of magnetization dynamics appeared to be the same within the excitation area for the uni- and bipolar IFE, the temporal period of the outgoing wavepackets decreases when the bipolar IFE is simulated.
\textcolor{newtext}{Individual $M_z$ time traces at different $x$ positions for all simulated cases are shown in Appendix~\ref{sec:appendixTimeTr}.}

We performed 1D FFT of $M_z$ temporal traces at each $x$ coordinate to show the evolution of the spectra when the MSSW wavepackets propagate outside the excitation area. 
The FFT spectra inside the central excitation spot at $x=0$ (red lines), between the spots at $x=1.5$\,$\mu$m (blue lines) and outside the excitation area at $x=10$\,$\mu$m (green lines) are shown for the CMA, uni- and bipolar IFE excitation in Fig.\,\ref{fig:FFT5pump}\,a,d and g, respectively.
These spectra are cross sections of the corresponding spatial evolution of the spectra (Fig.\,\ref{fig:FFT5pump}\,b,e and h) at certain coordinates $x$.
Also, in Fig.\,\ref{fig:FFT5pump}\,c,f and i, we show the respective dispersion characteristics of the excited MSSW obtained via 2D FFT of $M_z(t,x)$ maps, shown in Fig.\,\ref{fig:Mz5pump}\,b,d and f.
Analysis of these spectra allows us to identify the excited MSSW modes and analyze their spectral composition in detail.
In particular, all spectra have a quasi-discrete character (Fig.\ref{fig:FFT5pump}\,c,f and i). 
From a comparison of the calculated MSSW dispersion with the spectrum obtained from the analytical model Eq.~\eqref{FourierFull}, it is seen that the peaks in the spectra correspond to the predicted modes. 
In the case of CMA and unipolar IFE these are only even modes, while in the case of bipolar IFE -- only odd modes.
However, details of the spectra differ depending on the excitation mechanism.

Firstly, when the system is excited by the patterned CMA the most pronounced FFT amplitude around frequency \mbox{$f=4$\,GHz} appears to be localized between the excitation spots (blue line in Fig.\,\ref{fig:FFT5pump}\,a), and the MSSW with frequency above 4\,GHz is almost confined within the excitation area (black horizontal dashed lines in Fig.\,\ref{fig:FFT5pump}\,b).
At $x=0$ multiple separate peaks appear in the spectrum (red line in Fig.\,\ref{fig:FFT5pump}\,a) with the highest frequency around $f=4.5$\,GHz. 
Noticeably, the dynamics with such frequency is not found outside the excitation area.
The dynamics at $f=3.8,\,4,\,4.3$ and 4.5\,GHz corresponds to the MSSW modes of 2nd,\,4th,\,6th and 8th orders, respectively (Fig.\,\ref{fig:FFT5pump}\,c).
Also, the zero-order mode at $k=0$ has a broad spectral width, and the MSSW with a rather small $k$ are excited with FFT amplitude comparable to other modes, and propagates for the largest distance (Fig.\,\ref{fig:FFT5pump}\,b,\,c), as it has the highest group velocity.

Secondly, the unipolar IFE excitation results in the same set of even-order wavenumbers (Fig.\,\ref{fig:FFT5pump}\,f).
The main difference lies in the narrower total width of the dispersion and in the significant decrease of FFT amplitude of the zero-order mode, compared to the higher-order modes at any $x$.
The latter, as a result, leads to a decrease in the amplitude of the outgoing wavepacket compared to the second-order mode (Fig.\,\ref{fig:FFT5pump} e).
The reason of this is explained in terms of CMA and IFE temporal profiles, i.e. by the formation of the magnon trap, as discussed in the previous section.
Furthermore, under CMA excitation the low-frequency near-zero-$k$ mode is localized initally inside the resonator between the excitation spots at $x=1.5$\,$\mu$m (blue line in Fig.\,\ref{fig:FFT5pump}\,a and blue halos slightly above the corresponding horizontal black dashed line around $x=0$ and $k=0$ in Fig.\,\ref{fig:FFT5pump}\,b and c).
In turn, under IFE excitation, this mode is nearly absent within at the same position (Fig.\,\ref{fig:FFT5pump}\,d-f).
This confirms the possibility of inducing a magnon resonator when the CMA excitation is chosen instead of IFE.

In the case of bipolar IFE excitation, the spectra show the expected odd order MSSW modes (Fig.\,\ref{fig:FFT5pump}\,i).
Although low-frequency near-zero-$k$ mode is presented in the spectra at $x=0$, its amplitude is small (Fig.\,\ref{fig:FFT5pump}\,g), and  its propagation is not observed.
Instead, odd-order modes starting from the 1st propagate with lower group velocities compared to the unipolar IFE case, according to dispersion.
Thus, the choice of the IFE polarity allows one to control the order of the excited modes.

\section{Conclusion}

We presented a model for spin wave excitation, which describes the laser impact on the magnetic system in a single-pulse geometry and extended it to the case of periodically patterned Gaussian laser pulses.
The results demonstrate the way to excite multiple MSSW wavepackets propagating with different group velocities, whose dispersion corresponds to the set of selected quasi-discrete points.
The choice of the excitation mechanism, CMA or IFE, allows one to set whether the magnon resonator is induced and the localized modes are excited or not.
By considering the patterned SW source, we demonstrated that in reciprocal space it corresponds to the excitation of even or odd modes, the frequencies of which can be controlled by the polarity of the periodic impact and its spatial period.

The developed model proves and extends the flexible reconfigurability offered by optical methods of SW excitation.
Indeed, tunable periodical patterning of the spatial profile of the laser pulse can be achieved, for example, in the double pump geometry used for the creation of transient \textcolor{newtext}{gratings~\cite{januvsonis2016transient,Chang2018,carrara2022all}} or by a spatial light modulator~\cite{kolosvetov2022concept, Kolosvetov_IEEE2023}.
The model may find practical implementation in the development and modeling of complicated magnonic systems with multiple SW sources, which is one of the required conditions for the design of magnon logic devices~\cite{talmelli2020reconfigurable}.
Also, the presented approach may be useful to describe magnonic neuromorphic computing devices, as it offers a consideration of the system with a number of SW sources.
We note that our model can be readily improved to include in the consideration 2- and 3-dimensional spatially patterned SW sources and to excite and control other types of SW.
As for the next steps, we suggest the temporal modulation of the laser impact in magnonic systems, in addition to the spatial patterning considered here, as it makes possible to implement a comprehensive toolkit of spatio-temporal methods for tunable SW excitation and control of its parameters~\cite{jackl2017magnon,muralidhar2020sustained,hula2022spin}.
\textcolor{newtext}{Furthermore, including elastodynamics and spin-phonon interaction in the model~\cite{Deb2021,Azovtsev2023} would allow accounting for coherent acoustic waves driven by laser-induced thermal transient gratings that offer a promising way to control spin wave parameters due to magnetoelastic coupling~\cite{januvsonis2016transient,Chang2018,yaremkevich2023chip}.}

\section*{ACKNOWLEDGMENTS}
The work was supported by the Foundation for the Advancement of Theoretical Physics and Mathematics "BASIS" (grant No 22-1-5-122-1).

\section*{COMPETING INTERESTS}
The authors declare no competing interests.

\section*{author contributions}
\textbf{Iaroslav A. Filatov}: Conceptualization, Methodology, Investigation, Software, Writing -- Original draft preparation (equal), Writing -- Reviewing and Editing (equal).
\textbf{Petr I. Gerevenkov}: Data curation, Discussion (equal), Visualization, Writing -- Original draft preparation (equal).
\textbf{Nikolai E. Khokhlov}: Supervision (equal), Discussion (equal), Writing -- Reviewing and Editing (equal).
\textbf{Alexandra M. Kalashnikova}: Supervision (equal); Writing -- Reviewing and Editing (equal).

\section*{Data availability statement}
Data supporting the findings of this study are available from the corresponding author upon reasonable request.

\appendix

\section{Wavenumber distribution dependence on number of Gaussians $N$}\label{sec:appendixNgauss}
\begin{figure}[h]
    \centering
    \includegraphics{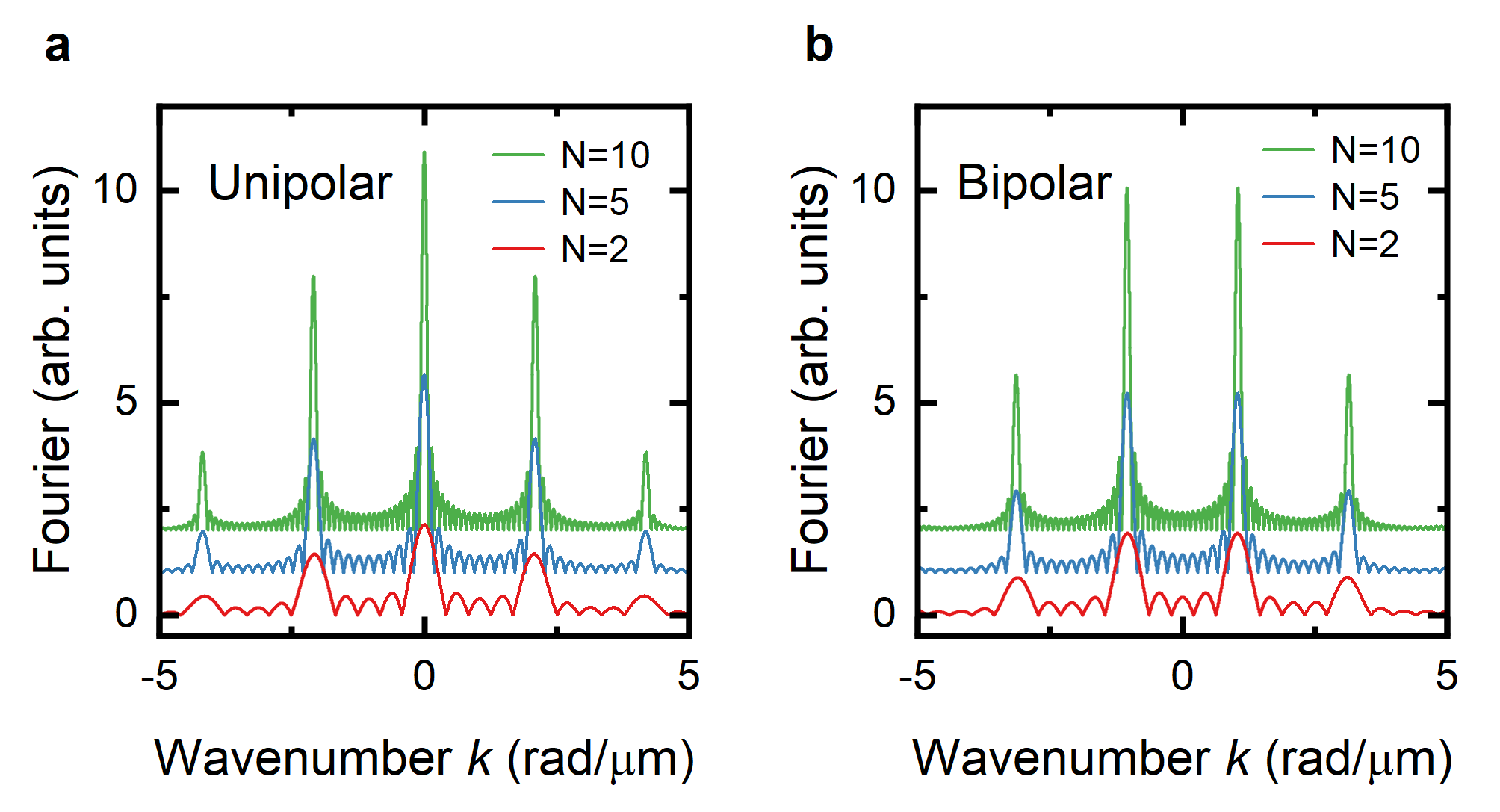}
    \caption{\textbf{Wavenumber distribution at different $N$}. \textbf{a} and \textbf{b}, uni- and bipolar cases, respectively.}
    \label{fig:Ngauss}
\end{figure}

Increase of $N$ does not lead to a change in the period and positions of the main peaks of the Fourier spectrum, while makes the main peaks narrower (Fig.~\ref{fig:Ngauss}). 

\begin{figure}[h]
    \centering
    \includegraphics{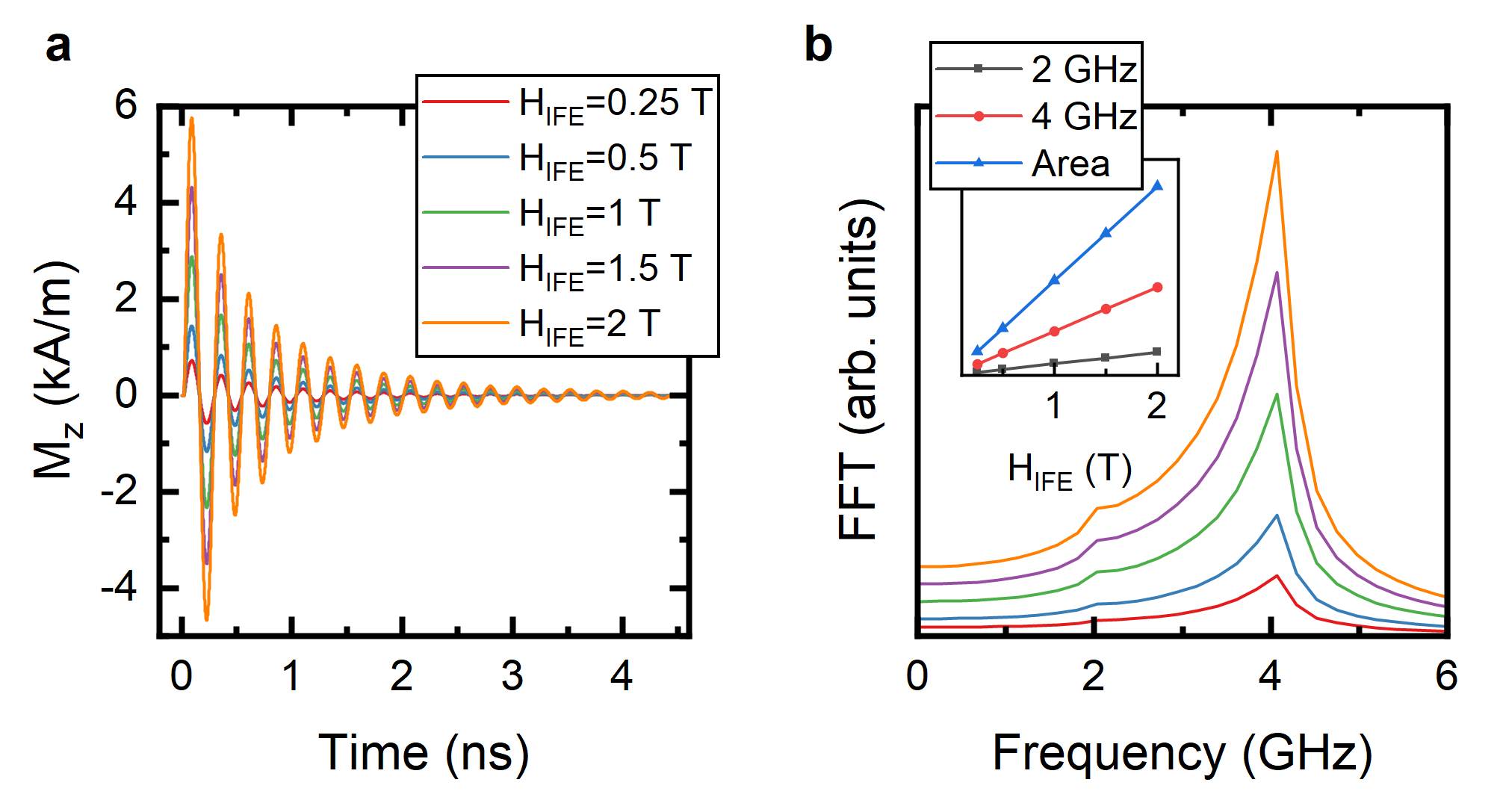}
    \caption{\textcolor{newtext}{\textbf{Magnetization dynamics under different IFE magnitudes upon single pulse excitation}. \textbf{a} Time traces $M_z(t,x=0)$ for $\mu_0 H_{\text{IFE}}$ in the range between 0.25 and 2\,T. \textbf{b}, FFT of signals from \textbf{a}. Inset in \textbf{b} shows the dependence of 2 and 4\,GHz frequency components amplitudes as well as the integral amplitude of the spectra on $H_{\text{IFE}}$.}}
    \label{fig:HIFEdep}
\end{figure}

\textcolor{newtext}{\section{Demonstration of linear regime of excitation}\label{sec:appendixHIFEdep}
In all the simulations using either IFE and CMA $M_z(t,x)$ doesn't exceed 3\,kA/m.
Using this value and $M_s$ we estimate the maximal deflection angle of magnetization, which has the value about $1^{\circ}$.
Thus, as the deflection angle of magnetization is rather small, the nonlinear effects do not not arise.
The absence of nonlinear effects is confirmed by the series of simulations using $\mu_0 H_{\text{IFE}}$ in the range between 0.25 and 2\,T, showing linear dependence of the spectral amplitudes on $H_{\text{IFE}}$ (Fig.~\ref{fig:HIFEdep}).}

\section{Spectral difference between CMA and IFE}\label{sec:appendixSpectral}

The Fourier image in the time domain of a femtosecond pulse with $\text{FWHM}_t = 200$\,fs has the form of a Gaussian function with $\sigma_f = \left( 2 \pi \sigma_t \right)^{-1}$ and the corresponding $\text{FWHM}_f \approx 4.4$\,THz. 
This corresponds to a close to uniform distribution of excited amplitudes over the entire GHz frequency range under consideration. 
Thus, the frequency spectrum of magnetization dynamics in the excitation area is determined by: (i) the Fourier image of excitation in space (see Sec.~\ref{subsec:k_distr}) and (ii) the density of states determined by the dispersion of magnetostatic waves (see Figs.~\ref{fig:singlepump}\,(d and h) and \ref{fig:FFT5pump}\,(c,f,i)) and inversely proportional to the group velocity. 
For a medium with selected parameters (see Sec.~\ref{subsec:modelParameters}) in the excited range of wave vectors, the dispersion exhibits a maximum of states near the frequency of 4\,GHz, which explains the predominance of this part of the spectrum over the FMR frequency of about 2\,GHz [see red lines in Figs.~\ref{fig:singlepump}\,(c and g) and \ref{fig:FFT5pump}\,(a,d,g)].

\section{Evolution of spin wave dynamics with distance}\label{sec:appendixTimeTr}

\begin{figure}[h]
    \centering
    \includegraphics{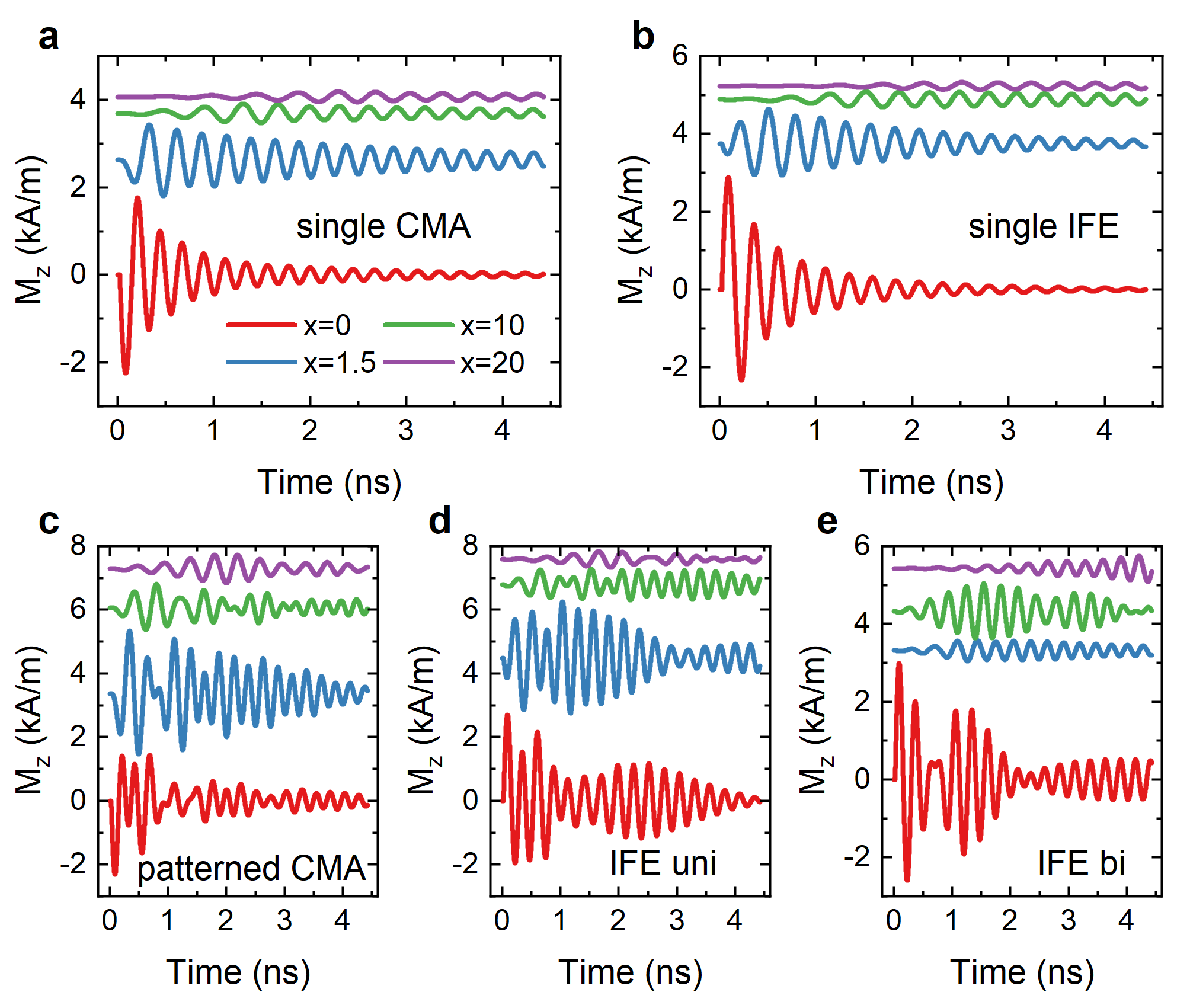}
    \caption{\textcolor{newtext}{\textbf{Magnetization dynamics at different spatial positions excited by single and periodic Gaussian impacts}. Time traces of magnetization dynamics at selected positions $x=0$, 1.5, 10 and 20\,$\mu$m. \textbf{a}, single pulse CMA. \textbf{b}, single pulse IFE. \textbf{c}, patterned CMA with 5 spots. \textbf{d}, unipolar IFE with 5 spots. \textbf{e}, bipolar IFE with 5 spots.}}
    \label{fig:timeTr}
\end{figure}

\bibliography{apssamp}

\end{document}